\documentclass[12pt,a4paper]{article}
\usepackage{graphicx,psfrag,amsmath,amssymb,amsfonts,latexsym,color,bm}
\usepackage[english]{babel}
\usepackage[utf8]{inputenc}
\usepackage{slashed}
\usepackage{physics}
\usepackage{tikz}
\usepackage{pgfplots}
\pgfplotsset{compat=1.17}
\usepackage[margin=2.5cm]{geometry}
\usepackage{simplewick}
\usepgfplotslibrary{polar}
\begin{document}
\title{On the spatial dependence of Casimir friction in graphene} 
\author{Aitor Fern\'andez and C\'esar D.~Fosco\\
{\normalsize\it Centro At\'omico Bariloche and Instituto Balseiro}\\
{\normalsize\it Comisi\'on Nacional de Energ\'{\i}a At\'omica}\\
{\normalsize\it R8402AGP S.\ C.\ de Bariloche, Argentina.} }
\maketitle
\begin{abstract}  
We study the spatial properties of the Casimir friction phenomenon for an
atom moving at a non-relativistic constant velocity parallel to a planar
graphene sheet.  The coupling of the atom to the vacuum electromagnetic
(EM) field is implemented by an electric dipole term, plus a R\"ontgen
term.  We study the fermion pair production, evaluating the angular
dependence of the fermion emission probability.
The phenomenon exhibits a threshold: it only exists when the speed of the
sliding motion is larger than the Fermi velocity of the medium.
\end{abstract}
\section{Introduction}\label{sec:intro} 
Quantum fluctuations produce macroscopic effects under the appropriate
circumstances, with the Casimir effect~\cite{Milonni:1994xx} being the most
celebrated example.   
A related kind of effect, also due to vacuum fluctuations,
may arise when the bodies move, or the boundary conditions they impose
become time-dependent. This can lead to dissipation, via real photons
excited out of the quantum vacuum, leading to what is known as the
dynamical Casimir effect~\cite{Dodonov:2020eto,diego_dalvit_DCE,Aubry:2021nnt}.
Yet another remarkable situation occurs when a purely quantum, dissipative
frictional force arises between bodies moving at a constant relative
velocity~\cite{Pendry_1997}. Here, the effect is due to the quantum degrees of freedom of the
media which are excited from the vacuum, and the EM field acting as
mediator. This Casimir friction effect has been studied extensively, though
some calculational issues have prompted debate~\cite{Pendry_2010_1,Leonhardt_2010,Pendry_2010_2}.

Here, we study this effect for an atom moving close to a graphene sheet,
evaluating the momentum distribution of the fermion pair which is created,
as a function of the parameters of the system. This study complements and
extends previous work~\cite{fosco2021motion_2} in two ways: the first is that, rather than evaluating the
total probability of vacuum decay, we focus on the angular aspects of the
phenomenon: how the probability of detecting the fermions on the plate
depends on the direction of emission, measured with respect to the trajectory of the
atom. The atom moves along a trajectory which is parallel to the graphene
plate, with a constant velocity. For the graphene system, we use its $2+1$
dimensional Dirac field description (see, for example~\cite{1111.3017v2,1608.03261v1}) , and the atom by an electron bounded to
the nucleus by a three-dimensional harmonic potential. 
The second way in which we introduce a novel ingredient is that the atom, in the
model we use, couples to the EM field through a dipole term, plus a
R\"ontgen term. The second term accounts for the fact that a moving
electric dipole carries also a magnetic dipole moment. This term can become
significant in certain scenarios, in particular in situations where there is 
quatum radiation, as shown in~\cite{neto_1996}.  
Quantum friction for two graphene sheets has been studied in~\cite{farias2017quantum};
note that in that situation there is no information (due to the geometry of the system) about
the spatial dependence of the pair production effect. Knowledge of that dependence
should, we believe, be relevant to the future design of nanodevices involving graphene.

The structure of this paper is as follows: in Sect.~\ref{sec:thesystem},
we describe the system and present the basic ingredients of our approach.
Then, in Sect.~\ref{sec:amplitudes}, we evaluate the probability
amplitudes for the relevant elementary process contributing to friction,
presenting a detailed study of its geometric (i.e., directional)
properties.
In Sect.~\ref{sec:conc} we present our conclusions.

\section{The system}\label{sec:thesystem}
The real-time action $\mathcal{S}$ for the system that we consider, may be
conveniently written as follows:
\begin{equation}\label{eq:defs}
\mathcal{S}[\bar{\psi},\psi,A,\mathbf{q};\mathbf{r}(t)] \;=\;
\mathcal{S}_\text{g}[\bar{\psi},\psi,A] \,+\,
\mathcal{S}_\text{a}[\mathbf{q}\hspace{2pt}] \,+\,\mathcal{S}_\text{em}[A]
\,+\, \mathcal{S}_{\text{a-em}}[A,\mathbf{q};\mathbf{r}(t)] \,,
\end{equation}
where $\mathcal{S}_\text{g}$ denotes a Dirac field action in $2+1$
dimensions, including its coupling to the EM field, while the terms
$\mathcal{S}_\text{a}$ and $\mathcal{S}_\text{em}$ denote the free
actions for the atom and the EM field, respectively.
$\mathcal{S}_\text{a-em}$ is the coupling between the atom and the EM
field.

It is worth pointing out the following: graphene is, in the continuum
version description which we are using here, described by a number $N$ of
flavours of $4$-component Dirac fields.  Each one of these flavours may be
thought of as composed of $2$ spinors transforming under an irreducible
representation of the $2+1$ dimensional Lorentz group, while the two
flavours are mixed by parity. We recall that, in $2+1$ dimensions, 
a parity transformation corresponds to a reflection, rather than a spatial
inversion (which is a rotation in $\pi$). For  the process we study here,
it will not make any difference which one of the $2 N$ $2$-component Dirac
fields is considered. Thus, we deal with just one of them and to find the
result in the general case one simply multiplies the result by $2 N$ (see
last paragraph of Sect.~\ref{sec:amplitudes}).

In this work, we adopt the following conventions: both $\hbar$ and the
speed of light are set equal to $1$, space-time coordinates are denoted by
$x^\mu$, $\mu\,=\, 0,\,1,\,2,\,3$, $x^0 = t$, and we use the Minkowski
metric \mbox{$g_{\mu\nu} \equiv {\rm diag}(1,-1,-1,-1)$}.  Dirac's
$\gamma$-matrices, on the other hand, are chosen to be in the
representation: $\gamma^0 \equiv \sigma_1$,  $\gamma^1 \equiv i \sigma_2$, 
$\gamma^2 \equiv i \sigma_3$,  where:
\begin{equation}\label{eq:gamma_matrices}
\sigma_1 \,=\, 
\left(
\begin{array}{cc}
	0 & 1 \\
	1 & 0
\end{array}
\right)
\;,\;\;
\sigma_2 \,=\, 
\left(
\begin{array}{cc}
	0 & -i\\
	i & 0 
\end{array}
\right) \;,
\end{equation}
and 
\begin{equation}
\sigma_3 \,=\,\left(
\begin{array}{cc}
	1 & 0 \\
	0 & -1
\end{array}
\right) \,,  
\end{equation}
where $\sigma_i$ ($i=1,\,2,\,3$) denote the usual Pauli's matrices.

Let us now describe the structure of each term in the action
(\ref{eq:defs}), beginning with the one corresponding to the atom: the
position of its center of mass, which to a very good approximation
coincides with that of its nucleus, is assumed to be externally driven, and
described by $\mathbf{r}(t) = (\mathbf{v} t,a)$. We have adopted a
reference system fixed to the graphene plane, which occupies the $x^3 = 0$
plane, $\mathbf{v}$ denotes the (constant) velocity of the atom, which
moves at a distance $a$ from the plate.
On the other hand, we assume that there is only one relevant (valence)
electron in the atom, and that its position with respect to its center of
mass is given by the vector: $\mathbf{q}$. In our description, therefore, the three
components of this vector are the only relevant degrees of freedom in the atom. 
Assuming that only single transitions are relevant to the process that we
are studying, the physics should be characterized by a single energy
(scale). 
it its sufficient to
take, as the  classical action accounting for the free dynamics of the
electron a harmonic one: 
\begin{equation}
\mathcal{S}_\text{a}[\mathbf{q}\hspace{2pt}]=\int dt
\left(\frac{1}{2}M\dot{\mathbf{q}}^{\hspace{4pt}2}-V(|\mathbf{q}|)\right)\approx
\int dt\frac{M}{2}\left(\dot{\mathbf{q}}^2-\Omega^2\mathbf{q}^2\right),
\end{equation}
where $M$ is the mass of the electron and $\Omega$ characterizes the
effective harmonic potential.

The free electromagnetic field has its dynamic given by the usual action, together with a gauge-fixing term
\begin{equation}\label{eq:accionA}
    \mathcal{S}_\text{em}[A]=\int d^4x~\left[-\frac{1}{4}F_{\mu\nu}F^{\mu\nu}-\frac{\lambda}{2}(\partial_\mu A^\mu)^2\right],
\end{equation}
where $F_{\mu\nu}=\partial_\mu A_\nu-\partial_\nu A_\mu$.

Graphene is a sheet of carbon atoms with a flat hexagonal crystal structure. This makes it effectively be described as a two-dimensional material. Furthermore, their electronic degrees of freedom can be described, at low energies, as Dirac fermions, and they satisfy a linear dispersion relation. That is, they behave like massless fermions that propagate with the Fermi velocity $v_F\approx0.003$~\cite{RevModPhys.81.109}. Its action is
\begin{equation}
    \mathcal{S}_\text{g}[\bar{\psi},\psi,A]=\int d^3x_\shortparallel~\bar{\psi}(x_\shortparallel)\Big(i\rho^\alpha_\beta\gamma^\beta D_\alpha-m\Big)\psi(x_\shortparallel),
\end{equation}
where $\rho^\alpha_\beta=\text{diag}(1,v_F,v_F)$ and $D_\alpha=\partial_\alpha+ie A_\alpha(x_\shortparallel,0)$. The solution to the free part takes the form
\begin{equation}
    \psi(x)=\sum_{s=\pm}\int\frac{d^2 \mathbf{p}}{2\pi}\sqrt{\frac{m}{p_0}}\Big(b(\mathbf{p},s)u(\mathbf{p},s)e^{-ip\cdot x}+d^\dagger(\mathbf{p},s)v(\mathbf{p},s)e^{ip\cdot x}\Big),
\end{equation}
where $u(\mathbf{p},s)$ and $v(\mathbf{p},s)$ satisfy
\begin{align}
    \sum_{s=\pm}u(\mathbf{p},s)\bar{u}(\mathbf{p},s)&=\frac{\rho^\alpha_\beta\gamma^\beta p_\alpha +m}{2m}\label{eq:uu}\\
    \sum_{s'=\pm}v(\mathbf{q},s')\bar{v}(\mathbf{q},s')&=\frac{\rho^\alpha_\beta\gamma^\beta q_\alpha -m}{2m}\label{eq:vv}.
\end{align}
$m$ is the mass gap parameter, which is almost zero for graphene. Taking the limit $m\to0$, the dispersion relation for fermions in graphene turns out to be $p_0\equiv E(\mathbf{p})=v_F\abs{\mathbf{p}}$.

On the other hand, the interaction action for the full system is 
\begin{align}\label{eq:interctionAction}
    \mathcal{S}_\text{int}[\bar{\psi},\psi,A,\mathbf{q};\mathbf{r}(t)]=&e\int d^4x~\Bigg[\rho^\sigma_\omega\hspace{1mm}\bar{\psi}(x_\shortparallel)\gamma^\omega\psi(x_\shortparallel)A_\sigma(x)\delta(x^3)+\nonumber\\
    &+\mathbf{q}\hspace{2pt}(x^0)\Big(\mathbf{E}(x)+\mathbf{v}\times \mathbf{B}(x)\Big)\delta^{(3)}(\mathbf{x}-\mathbf{r}(x^0))\Bigg].
\end{align}
The first term accounts for the coupling between the graphene, that lives in the plane $z\equiv x^3=0$, and the electromagnetic field, present in all space but evaluated in the plane of the graphene. The second term gives the coupling of the dipolar momentum of the atom, localized at $\mathbf{r}(t)$, with the electromagnetic field, having into account relativistic corrections up to order $\abs{\mathbf{v}}/c$ due to the movement of the dipole~\cite{Hnizdo_2012}.

\section{Probability amplitudes for quantum friction}\label{sec:amplitudes}
  The only role that the vacuum EM field plays in the processes that we
study is to mediate the excitations of the microscopic degrees of freedom
belonging to the two material objects involved in the phenomenon.
Thus, there will not be photons in the initial and final states.
Therefore, we shall consider the normalized initial ($\ket{\text{i}}$) and
final ($\ket{\text{f}}$) quantum states given by
\begin{align}
\ket{\text{i}}&=\ket{0_\text{a}}\otimes\ket{0_\text{em}}
\otimes\ket{0_\text {g}}\label{eq:inicialGrafeno}\\
\ket{\text{f}}&=\hat{a}^\dagger_i\ket{0_\text{a}}
\otimes\ket{0_\text{em}} \otimes\left(\frac{2\pi}{L}\right)^2
\hat{b}^\dagger(\mathbf{p},s)\hat{d}^\dagger(\mathbf{q},s')
\ket{0_\text{g}}\;,
\label{eq:finalGrafeno}
\end{align}
respectively.
That is, the system is initially at rest, while in the final one
the atom is in an excited state, corresponding to an electron excitation
for one of the three harmonic oscillator modes: the one in the direction
given by the index $i$. For graphene,  we assume a fermion-antifermion
pair; the fermion having  momentum $\mathbf{p}$ and spin $s$, while for
the antifermion those quantum numbers are $\mathbf{q}$ and $s'$.

For the states above, the first non trivial contribution to the amplitude
for the transition between them, appears to the second order in
the interaction action in (\ref{eq:interctionAction}), as it stems from
the usual perturbative expansion for the evolution operator.
Besides, for the states that we are using, the only non-vanishing
contractions, via Wick's theorem, follow from the `crossed' contributions.
Namely, contributions involving the coupling of the atom to the EM field
and also the interaction between the Dirac field and the EM field.

The resulting matrix element of the $S$-matrix to this order then becomes:
\begin{align}\label{eq:amTranGraf}
\mathcal{M}_i(\mathbf{p},\mathbf{q},s,s')=&\frac{i}{2!}\bra{\text{f}}
\mathbb{T}\left(\hat{S}_\text{int}^2\right)\ket{\text{i}}
=ie^2\bra{\text{f}} \hspace{-2pt}\int\hspace{-2pt}
d^4x\hspace{-2pt}\int\hspace{-2pt}
d^4y\hspace{2pt}\mathbb{T}\Big[\rho^\sigma_\omega
\bar{\psi}(x_\shortparallel)\gamma^\omega\psi(x_\shortparallel)A_\sigma(x)\delta(x^3)\times\nonumber\\
 &\times\mathbf{q}(y^0)\cdot\Big(\mathbf{E}(y)+\mathbf{v}\times
\mathbf{B}(y)\Big)\delta^{(3)}(\mathbf{y}-\mathbf{r}(y^0))\Big]\ket{\text{i
}}\; .
\end{align}
Wick's theorem also requires the knowledge of the contractions:
\begin{align}
 &\bcontraction{}{a}{{}_j} {q}a_j q_k(t)=\frac{1}{\sqrt{2M\Omega}}
e^{i\Omega t}\delta_{jk}\\
&\bcontraction{}{b}{(\mathbf{p},s)}{\bar\psi}b(\mathbf{p},s)
\bar\psi(x_\shortparallel)=
\frac{1}{2\pi}
\sqrt{\frac{m}{p_0}} e^{ip\cdot x_\shortparallel}~\bar{u}(\mathbf{p},s)\\
&\bcontraction{}{d}{(\mathbf{q},s')}{\psi}d(\mathbf{q},s')
\psi(x_\shortparallel)=\frac{1}{2\pi}\sqrt{\frac{m}{q_0}}
e^{iq\cdot x_\shortparallel}~v(\mathbf{q},s') \;.
\end{align}
We also need, as another ingredient to construct the amplitude, the
contraction between the gauge field $A_\sigma$
and $C_i \equiv (\mathbf{E}+\mathbf{v}\times\mathbf{B})_i$. This,
which can be computed by using the
free propagator of the gauge field, which in the Feynman gauge is:
\begin{equation}\label{eq:propA}    
\bcontraction{}{A}{{}_\mu(x)}{A}A_\mu(x)A_\nu(y)=G_{\mu\nu}(x-y)=\int\frac{
d^4k}{(2\pi)^4}\frac{-i\hspace{2pt}g_{\mu\nu}}{k^2+i\epsilon}e^{-ik\cdot(x-
y)} \;.
\end{equation}

Putting together the previous elements, after a lengthy but
otherwise straightforward calculation we find that the transition
amplitude becomes:
\begin{equation}
    \mathcal{M}_i(\mathbf{p},\mathbf{q},s,s')
= \delta\Big(\chi(\mathbf{p},\mathbf{q})\Big) K^\sigma I_{\sigma i}(p+q)
\;,
\end{equation}
where we have introduced
\begin{equation}\label{eq:K}
    K^\sigma=\left(\frac{2\pi}{L}\right)^2\frac{ie^2m}{(2\pi)^2
\sqrt{2M\Omega p_0 q_0}}\rho^\sigma_\omega
\bar{u}(\mathbf{p},s)\gamma^\omega v(\mathbf{q},s') \;,
\end{equation}
\begin{equation}\label{eq:Isigma}
   I_{\sigma i}(p+q)=\pi e^{-a\sqrt{-(p+q)^2}}\times\left\{
  \begin{array}{ll}
       \displaystyle \frac{1}{\sqrt{-(p+q)^2}}\Big(\Omega \eta_{\sigma i}+(p+q)_i(\eta_{\sigma 0}+v_\sigma)\Big)  & \text{for  } i=1,2 \\
       & \\
   i(\eta_{\sigma 0}+v_\sigma) & \text{for  } i=3
\end{array}
\right.
\end{equation}
and
\begin{equation}\label{eq:chiDefinition}
\chi(\mathbf{p},\mathbf{q})\equiv\Omega+v_F(\abs{\mathbf{p}}+\abs{\mathbf{q
}})-(\mathbf{p}+\mathbf{q})\cdot \mathbf{v} \;.
\end{equation}
Note that the last object, being the argument of a  Dirac's $\delta$
function, provides important kinematic information about
the process. Firstly, the {\em total\/} momentum that appears in graphene
as a consequence of the created pair, has a positive component in the
direction of the atom's velocity. Another observation is that, for
this process to happen, the velocity of the atom should be greater than
Fermi's velocity in graphene, i.e., $\abs{\mathbf{v}}>v_F$. Besides,
$(p+q)^2<0$, so that $p+q$ must be a space-like momentum. In other words,
the Coulombian part of the EM interaction is prevalent.
At this point, it is worth mentioning some relevant observations regarding
Lorentz invariance. It is well-known that fermions in graphene have a
`relativistic' dispersion relation, with $v_F$ playing the role of
the speed of light, and with spacetime reduced to $2+1$ dimensions. They
 behave like massless particles moving with a velocity $v_F$ on the
plane. Since $v_F$  is less than the speed of light in the vacuum, they
can have a total space-like momenta without involving non-physical
superluminical particles. That reconciles the  fact that the momentum
$p+q$ is space-like  with the creation of real particles.
Furthermore, $(p+q)^2<0$ is consistent with the fact that the final state
contains no real photons.

The transition we have up to now, corresponds to a final state in
which the spin of the fermions, their momenta and the orientation of the
excitation of the atom have a specific value. One is usually interested
in the knowledge of the probability as a function of momentum, regardless
of the spin of the fermions, and of the direction of the harmonic
excitation on the electron in the atom. This  amounts to adding the
probability densities
$|\mathcal{M}_i(\mathbf{p},\mathbf{q},s,s')|^2$
for every value of $s$ and orientation $i$.
The resulting probability per unit time is then a function of the two
momenta $\mathbf{p}$ and $\mathbf{q}$:
\begin{equation}
    \mathcal{P}(\mathbf{p},\mathbf{q})
=\frac{1}{T}\sum_{s,s'}\sum_{i=1}^3
\abs{\mathcal{M}_i(\mathbf{p},\mathbf{q} ,s,s')}^2 \;.
\end{equation}
The sum over spins and oscillator directions allows one to produce a more
explicit expression for that probability per unit time. Indeed, by using
(\ref{eq:uu}), (\ref{eq:vv})  plus the $2+1$ dimensional Dirac matrices
trace relation:
\begin{equation}
\tr{\gamma^\sigma\gamma^\mu\gamma^\lambda\gamma^\nu}
= 2(\eta^{\sigma\mu}\eta ^{\lambda\nu}
-\eta^{\sigma\lambda}\eta^{\mu\nu}+\eta^{\sigma\nu}
\eta^{\lambda\mu}) \;,
\end{equation}
the result for ${\mathcal P}$ may be put in the form:
\begin{equation}
\mathcal{P}(\mathbf{p},\mathbf{q};\Omega,a)=
\frac{e^{-2a\sqrt{-(p+q)^2}}}{\Omega v_F^2\abs{\mathbf{p}}\abs{\mathbf{q}}}
\delta\Big(\chi(\mathbf{p},\mathbf{q}
;\Omega)\Big)F(\mathbf{p},\mathbf{q};\Omega) \;,
\end{equation}
with
\begin{align}\label{eq:F}
F(\mathbf{p},\mathbf{q};\Omega)
=&
\frac{1}{-(p+q)^2}\Bigg\{\Big(\abs{\mathbf{p}+\mathbf{q}}^2-(p+q)^2\Big)
\Big((p_0-v_F^2\mathbf{p}\cdot \mathbf{v})(q_0-v_F^2\mathbf{q}\cdot
\mathbf{v})+\nonumber\\
&-(1-v_F^2 v^2)(p_0 q_0 -v_F^2\mathbf{p}\cdot\mathbf{q})/2\Big)+\Omega^2
v_F^2 p_0 q_0+\\
&+\Omega
v_F^2(\mathbf{p}+\mathbf{q})\cdot\Big((q_0-v_F^2\mathbf{q}\cdot\mathbf{v})
\mathbf{p}+(p_0-v_F^2\mathbf{p}\cdot\mathbf{v})\mathbf{q}-(p_0
q_0-v_F^2 \mathbf{p}\cdot\mathbf{q})\mathbf{v}\Big)\Bigg\}\Bigg|_\text{on
shell}\hspace{-8mm}\nonumber \;.
\end{align}
We have used the `on shell' expression to mean that the fermion satisfy
the dispersion relations of real particles in graphene:
$p_0=v_F\abs{\mathbf{p}}$ and $q_0=v_F\abs{\mathbf{q}}$. In order to
further clarify the dependence of the result on all the relevant parameters
of the model, we have also made explicit the dependence on the dimensional
parameters $a$ and $\Omega$.

In spite of its cumbersome appearance, it is not difficult to verify (as a
consistency check) analytically that $F(\mathbf{p},\mathbf{q};\Omega)\geq 0$ 
when the R\"ontgen term coupling is ignored. One should use, in order to
verify that inequality, the relation $\Omega=-v_F(\abs{\mathbf{p}}+\abs{\mathbf{q}})$.
With the R\"ontgen term included, we have verified numerically that
$F(\mathbf{p},\mathbf{q};\Omega)\geq 0$.

In order to have a more explicit knowledge of the angular dependence of the
effect, we begin by introducing modules and angles for the relevant
vectors. First, we note that
\begin{equation}
     F(\mathbf{p},\mathbf{q};\Omega)=v_F^2\abs{\mathbf{p}}\abs{\mathbf{q}}f(\mathbf{p},\mathbf{q};\Omega),
\end{equation}
so that
\begin{equation}
\mathcal{P}(\mathbf{p},\mathbf{q};\Omega,a) =
\Omega^{-1}\delta\Big(\chi(\mathbf{p},\mathbf{q};\Omega)\Big)
\underbrace{e^{-2a\sqrt{-(p+q)^2}}
f(\mathbf{p},\mathbf{q};\Omega)}_{g(\mathbf{p},\mathbf{q};\Omega,a)}
\end{equation}
and
\begin{equation}
\chi(\mathbf{p},\mathbf{q};\Omega)=
\Omega+\abs{\mathbf{p}}(v_F-v\cos\theta_p)+\abs{\mathbf{q}}(v_F-v\cos\theta_q)=0
\;.
\end{equation}
The $\delta$-function may then be used to fix the value of
$\abs{\mathbf{q}}$, since:
\begin{equation}\label{eq:simpleDelta}
\delta \left(\chi(\mathbf{p},\mathbf{q};\Omega)\right)=
\frac{\delta\Big(\abs{\mathbf{q}}-q_0(\mathbf{p},\theta_q;\Omega)\Big)}{\abs{v_F-v\cos\theta_q}}
\;,
\end{equation}
where we have introduced:
\begin{equation}
q_0(\mathbf{p},\theta_q;\Omega)=
\frac{\Omega+\abs{\mathbf{p}}(v_F-v\cos\theta_p)}{v\cos\theta_q-v_F} \equiv 
\frac{s(\mathbf{p};\Omega)}{v\cos\theta_q-v_F} \;.
\end{equation}

Just positive values of $q_0$ are physical. Imposing
$q_0(\mathbf{p},\theta_q;\Omega)>0$ determined the allowed region $\mathcal{R}$ 
for the angle between $\mathbf{q}$ and $\mathbf{v}$. Defining
$\cos\alpha \,\equiv\, v_F/v$, $\mathcal{R}$ is given by:
\begin{equation}\label{eq:region}
    \theta_q \in \left\{ 
    \begin{array}{lc}
       {[}0,\alpha)\cup(2\pi-\alpha,2\pi)  & \text{if  } s(\mathbf{p};\Omega)>0 \\
       & \\
       (\alpha,2\pi-\alpha) &  \text{if 
 } s(\mathbf{p};\Omega)<0 \;.
    \end{array}
    \right.
\end{equation}

For the probability of detecting any given particle of the pair, with
momentum $\mathbf{p}$, we compute (note that the probability is
symmetric under the exchange of particle by antiparticle):
\begin{equation}
\mathcal{P}(\mathbf{p};\Omega,a)=\int
	d^2\mathbf{q}~\mathcal{P}(\mathbf{p},\mathbf{q};\Omega,a)=\int_\mathcal{R}d\theta_q\int\limits_0^\infty
d\abs{\mathbf{q}}~\abs{\mathbf{q}}\mathcal{P}(\mathbf{p},\abs{\mathbf{q}},
\theta_q;\Omega,a)
\;.
\end{equation}
Taking into account (\ref{eq:region}), the angular integral becomes:
\begin{equation}
\int_\mathcal{R}
d\theta_q=\Theta\left[s(\mathbf{p};\Omega)\right]\left(\int\limits_{0}^{\alpha}d\theta_q\hspace{1mm}+\int\limits_{2\pi-\alpha}^{2\pi}\hspace{-1mm}d\theta_q\right)\hspace{1mm}+\hspace{2mm}\Theta\left[-s(\mathbf{p};\Omega)\right]\hspace{-2mm}\int\limits_{\alpha}^{2\pi-\alpha}\hspace{-2mm}d\theta_q
\,.
\end{equation}
By using (\ref{eq:simpleDelta}), we obtain
\begin{align}\label{eq:probP}
\mathcal{P}(\mathbf{p};\Omega,a)=&\Omega^{-1}
s(\mathbf{p};\Omega)\Bigg\{\Theta\left[s(\mathbf{p};\Omega)\right]
\left(\int\limits_{0}^{\alpha}d\theta_q\hspace{1mm} + 
\int\limits_{2\pi-\alpha}^{2\pi}\hspace{-1mm}d\theta_q\right)-\hspace{2mm}
\Theta\left[-s(\mathbf{p};\Omega)\right]\hspace{-2mm}\int\limits_{\alpha}^{2\pi-\alpha}\hspace{-2mm}d\theta_q\hspace{1mm}\Bigg\}\times\nonumber\\
&\times
\frac{g\big(\mathbf{p},q_0(\mathbf{p},\theta_q;\Omega),\theta_q;\Omega,a\big)}{\abs{v\cos\theta_q-v_F}^2}=
\mathcal{P}\left(\mathbf{p}/\Omega;1,a\Omega\right) \,.
\end{align}
The last equality above follows from the homogeneity properties of the
functions involved. This will allow us to get expressions and plots, in
terms of fewer parameters than one might have expected a priori.

An exact relation that one can see from the previous expression corresponds
to finding the angle for which the probability vanishes, what sets the
angular width for the production of pairs. It follows from the observation
that:
\begin{equation}
s(\abs{\mathbf{p}},\theta_p^0;\Omega)=0\hspace{3mm}\Rightarrow\hspace{3mm}\theta_p^0=\arccos\left[\frac{1}{v}\left(v_F+\frac{\Omega}{\abs{\mathbf{p}}}\right)\right]\stackrel{\abs{\mathbf{p}}\gg\Omega}{\longrightarrow}\alpha
\;.
\end{equation}  
This also implies $\abs{\mathbf{q}}=0$, and the probability (\ref{eq:probP}) vanishes.

Another probability per unit time follows by just asking for the
probability density per unit angle of detecting any particle, regardless of
their momentum: 
\begin{equation}
\mathcal{P}(\theta_p;\Omega,a)=\int\limits_0^\infty
	d\abs{\mathbf{p}}~\abs{\mathbf{p}}\mathcal{P}(\mathbf{p};\Omega,a)=\Omega^2\mathcal{P}(\theta_p;1,a\Omega)
\;.
\end{equation}
This distribution has been plotted in Fig.~\ref{fig:polar}. 

We see that, for velocities close to $v_F$, the probability becomes highly
concentrated around the direction along which the atom moves. On the other
hand, it widens up as the velocity increases. The area inside each
curve is proportional to the total probability of this process to happen,
and we see that it reaches its maximum around $v=4.5\cdot10^{-3}\sim
1.5v_F$.

\pgfplotstableread{graficoPolarDef.dat}{\loadeddata}
\begin{figure}
    \centering
    \begin{tikzpicture}
        \begin{polaraxis}[
            legend style={at={(0.5,-0.3)}},
            rotate=90,
            xticklabel style={anchor=-\tick-90},
            minor grid style={dotted},
            minor x tick num=1,
            minor y tick num=1,
            xtick={0,30,...,330},
            yticklabels={null}
            ]
            \addplot[
                data cs=polar,
                ultra thick,
                red,
                ] table[x index=0,y index=1] {\loadeddata};
            \addlegendentry{$v=3.4\cdot 10^{-3}$},
            \addplot[
                data cs=polar,
                ultra thick,
                orange,
                ] table[x index=0,y index=2] {\loadeddata};
            \addlegendentry{$v=3.6\cdot 10^{-3}$},
            \addplot[
                data cs=polar,
                ultra thick,
                yellow,
                ] table[x index=0,y index=3] {\loadeddata};
            \addlegendentry{$v=3.8\cdot 10^{-3}$},
            \addplot[
                data cs=polar,
                ultra thick,
                green,
                ] table[x index=0,y index=4] {\loadeddata};
            \addlegendentry{$v=4.0\cdot 10^{-3}$},
            \addplot[
                data cs=polar,
                ultra thick,
                blue,
                ] table[x index=0,y index=5] {\loadeddata};
            \addlegendentry{$v=4.5\cdot 10^{-3}$},
            \addplot[
                data cs=polar,
                ultra thick,
                magenta,
                ] table[x index=0,y index=6] {\loadeddata};
            \addlegendentry{$v=5.5\cdot 10^{-3}$},
            \addplot[
                data cs=polar,
                ultra thick,
                black,
                ] table[x index=0,y index=7] {\loadeddata};
            \addlegendentry{$v=8.0\cdot 10^{-3}$},
        \end{polaraxis}
\end{tikzpicture}
\caption{Polar distribution of the probability for different velocities
of the atom.}
    \label{fig:polar}
\end{figure}
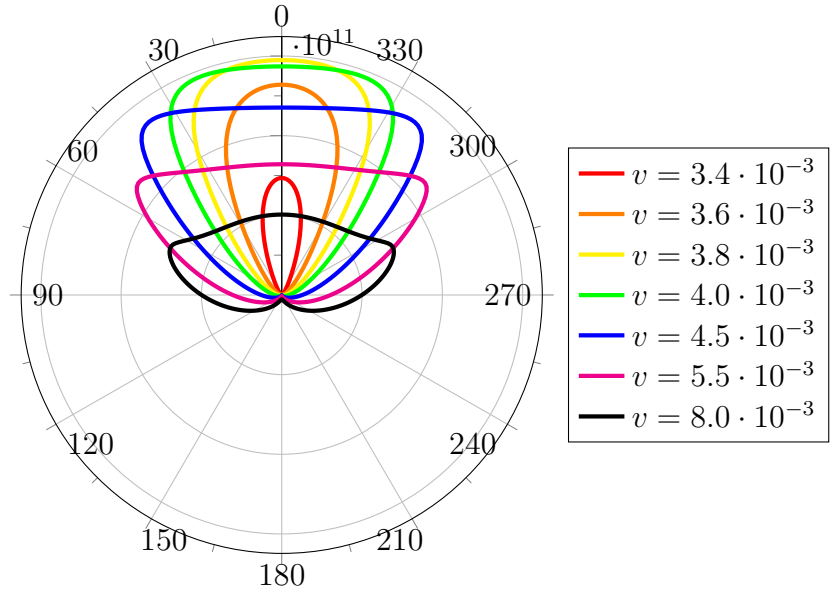

Another interesting quantity is the power dissipated $\mathcal{W}$, which
is related to the friction force by $\mathcal{W}=v F_\text{fr}$. The
dissipated power is the energy per unit time transmitted from the
mechanical system that moves the atom to the graphene through the
electromagnetic field. The energy that the graphene is receiving when a
fermionic pair of momenta $\mathbf{p}$ and $\mathbf{q}$ is created is
$E=v_F(\abs{\mathbf{p}}+\abs{\mathbf{q}})$, so the power transmitted to 
graphene is proportional to
\begin{align}
    \mathcal{W}(\Omega,a)=&\int d^2\mathbf{p}\int d^2\mathbf{q}~(\abs{\mathbf{p}}+\abs{\mathbf{q}})\mathcal{P}(\mathbf{p},\mathbf{q};\Omega,a)\\
    =&\int\limits_0^{2\pi}d\theta_p\int\limits_0^\infty d\abs{\mathbf{p}}\abs{\mathbf{p}}s(\mathbf{p};\Omega)\Bigg\{\Theta\left[s(\mathbf{p};\Omega)\right]\left(\int\limits_{0}^{\alpha}d\theta_q\hspace{1mm}+\int\limits_{2\pi-\alpha}^{2\pi}\hspace{-1mm}d\theta_q\right)-\hspace{2mm}\Theta\left[-s(\mathbf{p};\Omega)\right]\hspace{-2mm}\int\limits_{\alpha}^{2\pi-\alpha}\hspace{-2mm}d\theta_q\hspace{1mm}\Bigg\}\times\nonumber\\
    &\times \left(\abs{\mathbf{p}}+\frac{s(\mathbf{p})}{v\cos\theta_q-v_F}\right)\frac{g\big(\mathbf{p},q_0(\mathbf{p},\theta_q;\Omega),\theta_q;\Omega,a\big)}{\abs{v\cos\theta_q-v_F}^2}=\Omega^3\mathcal{W}(1,a\Omega)
\end{align}
We have not written it explicitly but this is a function of the velocity of the atom (appearing also in $\alpha,s(\cdot\cdot\cdot)$ and $g(\cdot\cdot\cdot)$), so we could see the dependence of the friction force on the velocity.

All our calculations have been made for a $2$-component Dirac field.
Graphene, however, does correspond to $N$ flavours of $4$-component Dirac
fields, each one of those $N$ flavours in a reducible representation of the Lorentz
group in $2+1$ dimensions. The coupling between each flavour and the EM
field is the same; therefore, in this kind of process, all the results for
the probabilities should be multiplied by a factor $2 N$ (we assume one is
interested in the probability of creating fermions of {\em any\/} flavour).

\section{Conclusions}\label{sec:conc}
We have presented a detailed calculation of the process that drives Casimir
Friction when an atom moves close to a graphene plate, presenting the
angular dependence of the probability of detecting fermions, as a function
of the parameters of the system. All our calculations have been presented
for a single $2$-component Dirac flavour. Results corresponding to 
$N$ $4$-component Dirac flavours can be obtained by multiplying our results
for the probability by a global factor $4 N$. 

Besides the known fact that there is a velocity threshold for the effect to
occur, we have found a relation which restricts, for particles with a given momentum,
the angular region where they could be detected.

We have also obtained the angular probability distribution, namely, the
probability density (per unit time) of detecting a given particle
regardless of its momentum.

A related but different observable is the power dissipation on the graphene
plate, for which we could obtain expressions depending on the velocity $v$
and the other parameters $a$, $\Omega$, $v_F$. 

We think it is worth mentioning the following observation: since the fermion and
antifermion have opposite electric charges, and the probability of
detecting a fermion is identical to the one of detecting an antifermion
(for the same momenta), the processes described here do not amount to the
production of a net electric current. 
However, we suggest that a possible way to allow for the production of a net
current on the sample would be to study friction in the presence of a
constant and uniform magnetic field, normal to the graphene plane. Under
this external condition, particle and antiparticle will experience opposite
forces when they are produced along a given direction, with the same
velocity.

\section*{Acknowledgements}
This work was supported by ANPCyT, CONICET, UBA and UNCuyo.
\bibliographystyle{unsrt}
\bibliography{references.bib}

\begin{thebibliography}{10}

\bibitem{Milonni:1994xx}
P.~W. Milonni.
\newblock {\em {The Quantum vacuum: An Introduction to quantum
  electrodynamics}}.
\newblock Academic Press, 1994.

\bibitem{Dodonov:2020eto}
V.~V. Dodonov.
\newblock {Fifty Years of the Dynamical Casimir Effect}.
\newblock {\em MDPI Physics}, 2(1):67–104, 2020.

\bibitem{diego_dalvit_DCE}
Diego A.~R. Dalvit, Paulo A.~Maia Neto, and Francisco~Diego Mazzitelli.
\newblock {\em {Fluctuations, Dissipation and the Dynamical Casimir Effect}}.
\newblock Springer, 2011.

\bibitem{Aubry:2021nnt}
B.~A.~Juárez Aubry and R.~Weder.
\newblock {A short review of the Casimir effect with emphasis on dynamical
  boundary conditions}.
\newblock {\em Rev. Mex. Fis. Suppl.}, 3(2):020714, 2022.

\bibitem{Pendry_1997}
J.~B. Pendry.
\newblock {Shearing the vacuum - quantum friction}.
\newblock {\em Journal of Physics Condensed Matter}, 9(47):10301–10320,
  November 1997.

\bibitem{Pendry_2010_1}
J.~B. Pendry.
\newblock {Quantum friction-fact or fiction?}
\newblock {\em New Journal of Physics}, 12(3):033028, March 2010.

\bibitem{Leonhardt_2010}
U.~Leonhardt.
\newblock {Comment on 'Quantum Friction—Fact or Fiction?'}.
\newblock {\em New Journal of Physics}, 12(6):068001, June 2010.

\bibitem{Pendry_2010_2}
J.~B. Pendry.
\newblock {Reply to comment on 'Quantum friction—fact or fiction?'}.
\newblock {\em New Journal of Physics}, 12(6):068002, June 2010.

\bibitem{fosco2021motion_2}
C.~D. Fosco, F.~C. Lombardo, and F.~D. Mazzitelli.
\newblock {Motion-Induced Radiation Due to an Atom in the Presence of a
  Graphene Plane. Universe 2021, 7, 158}, 2021.

\bibitem{1111.3017v2}
I.~V. Fialkovsky and D.~V. Vassilevich.
\newblock {Quantum Field Theory in Graphene}.
\newblock {\em Int. J. Mod. Phys. A}, 27:1260007, November 2012.

\bibitem{1608.03261v1}
Ignat~V. Fialkovsky and Dmitri~V. Vassilevich.
\newblock {Graphene through the looking glass of QFT}.
\newblock {\em Mod. Phys. Lett. A}, 31(40), August 2016.

\bibitem{neto_1996}
P.~A.~Maia Neto and L.~A.~S. Machado.
\newblock {Quantum radiation generated by a moving mirror in free space}.
\newblock {\em Physical Review A}, 54(4):3420, 1996.

\bibitem{farias2017quantum}
M.~B. Farias, C.~D. Fosco, F.~C. Lombardo, and F.~D. Mazzitelli.
\newblock {Quantum friction between graphene sheets}.
\newblock {\em Physical Review D}, 95(6):065012, 2017.

\bibitem{RevModPhys.81.109}
A.~H. Castro~Neto, F.~Guinea, N.~M.~R. Peres, K.~S. Novoselov, and A.~K. Geim.
\newblock The electronic properties of graphene.
\newblock {\em Rev. Mod. Phys.}, 81:109--162, Jan 2009.

\bibitem{Hnizdo_2012}
V~Hnizdo.
\newblock Comment on ‘electromagnetic force on a moving dipole’.
\newblock {\em European Journal of Physics}, 33(1):L3, nov 2011.

\end{thebibliography}
\end{document}